\documentclass[12pt,aps]{iopart}

\usepackage{graphicx}
\usepackage{dcolumn}
\usepackage{bm}

\begin{document}

\title[Shape modulated MC]{Magnonic band gap design by the edge modulation of micro-sized waveguides}

\author{F Ciubotaru$^{1}$, A V Chumak$^{1}$, N Yu Grigoryeva$^{2}$, A A Serga$^{1}$, and B Hillebrands$^1$}

\address{$^1$ Fachbereich Physik and Forschungszentrum OPTIMAS, Technische Universit\"{a}t Kaiserslautern, 67663 Kaiserslautern,
 Germany}

\address{$^2$ Department of Physical Electronics and Technology, St. Petersburg Electrotechnical University, 197376
Saint Petersburg, Russia}

\ead{ciubotaru@physik.uni-kl.de}

\begin{abstract}
The potential to control the number of the spin-wave band gaps of
a magnonic crystal (MC) by variation of its geometry is
investigated by numerical simulations. The magnonic crystal is
represented by a micro-sized planar ferromagnetic waveguide with
periodically variable width. By choosing a step-like or sinusoidal
variation of the width, the magnonic crystal reveals multiple or
single band gaps, respectively. This allows for additional degrees
of freedom in the design of MC-based microwave filters and phase
shifters with desired characteristics. The MCs band gaps have been
studied in the space and frequency domains exploring the spin-wave
spectrum dependence on the probing position inside the magnonic
crystal.
\end{abstract}


\submitto{\JPD}

\maketitle

The linear and nonlinear magnetization dynamics in magnonic
crystals (MCs) \cite{Nikitov, APL-chumak, JAP-chumak, Krawczyk,
APLChumak, Kalinsoliton, JPDChumak, GubbiottiAPL, Kostylev, Wan09,
NatCom, Dvornik} attracts increasing attention due to the
potential application of these devices for signal processing in
the microwave frequency range \cite{Wan09, KimHan, KimLee, Krug,
Chumak-muMC}. Significant progress has recently been made in the
development of magnonic crystals based on yttrium-iron-garnet
(YIG) ferrite  films \cite{Nikitov, APL-chumak, APLChumak,
Kalinsoliton, JPDChumak, NatCom, ChiZhu} having extremely small
spin-wave damping \cite{Gurevich, YIG-magnonics}. The attention
has been focused on two types of YIG based magnonic crystals: (1)
crystals designed as spin-wave waveguides of periodically variable
thickness \cite{APL-chumak, JAP-chumak} and (2) crystals formed by
placing of a plane spin-wave waveguide in a bias magnetic field
spatially modulated by electrical currents \cite{JPDChumak}.
Grooved structures represent probably the simplest design of an
one-dimensional magnonic crystal, while the current controlled MCs
propose such unique features as gradual tuning and fast
manipulation of crystal properties \cite{NatCom,
karenowska-PRL-2011}.

The transmission characteristics of these two types of MCs are
markedly different: whereas a current-controlled MC has only one
rejection-band (i.e. the frequency region where spin waves are not
allowed to propagate), a grooved MC possesses multiple
rejection-bands in its transmission band. This difference is
clarified if one takes into account that the reflection
coefficient of a wave with wavenumber $k$ is proportional to the
$2k$ Fourier component of the magnetic periodic structure
\cite{JPDChumak}. The spatial variations of the magnetic
properties of the grooved crystals can be described by a
\emph{step-like function} expressing the change of the YIG film
thickness with a quasi-rectangular profile \cite{APL-chumak}. The
current-controlled  crystal is designed in such a way that
variations of the bias magnetic field are practically
\emph{sinusoidal} \cite{JPDChumak}. Since the spectrum of the
latter crystal contains only one space component, the single
rejection-band appears solely.

Thus, by choosing a profile of a magnetic non-uniformity in the
YIG film one can control not only the width and depth of the
rejection bands \cite{JAP-chumak} but also their number. It
allows, for example, creating of single- and multiple-band
microwave filters. However, YIG-based structures do not satisfy
the demands of modern microwave technology mainly due to their
comparatively large sizes. Therefore, it opens up the necessity to
design planar MCs of micron and sub-micron size with desired band
gaps characteristics.

Here we propose a method to control the transmission band gap
characteristics of a microstructured magnonic crystal. Using
micromagnetic simulations \cite{OOMMF} we analyze and compare two
magnonic crystals designed as Permalloy (Ni$_{80}$Fe$_{20}$, Py)
spin-wave waveguides of periodically variable width. The
structures are planar and can easily be integrated on a chip
within a CMOS process. The width of one waveguide is according to
a step function (that forms rectangular notches), while the width
of another structure follows a sinusoidal function. We show that
by substituting the sharp notches with a harmonic edge profile one
can change the MC operational transmission characteristics from a
multiple rejection band state to a single band state. The role of
spin-wave channeling, spin-wave wavelength transformation, and
geometrical scattering in the formation of the crystal
characteristics is discussed.

\begin{figure}
\includegraphics[width=9.5 cm]{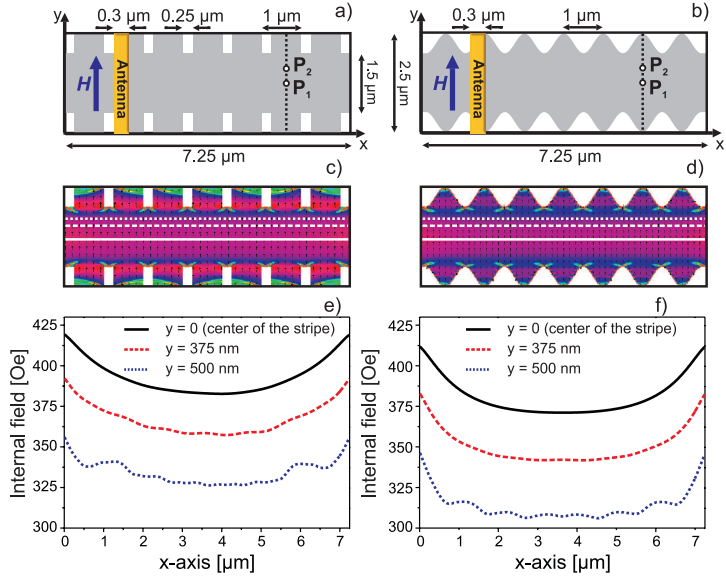}
\caption{\label{Fig1} (Color online) Sketch of the simulated
structures: MCs with rectangular a) and sinusoidal notches b). c)
and d) Distribution of the magnetic internal field ($H_{y}$
component) for the simulated structures. e) and f) $H_{y}$
component of the internal field extracted along the magnetic
stripes in the center (y = 0, white lines from (a) and (b)) at
half the distance between the center and the edge of the notches
(y = 375 nm, dashed line), and at y = 500 nm (dotted blue line).}
\end{figure}

The two types of magnonic crystals under investigation are
presented in Fig.~\ref{Fig1}a and Fig.~\ref{Fig1}b. They consist
of 7.25~$\mu$m-long Permalloy waveguides of 40~nm thickness. The
width of both MCs varies between $w_{0}=2.5~\mu$m and
$w_{1}=1.5~\mu$m with a period $\Lambda= 1~\mu$m. The geometry was
chosen in such a way that the structure with rectangular notches
is similar to the MC studied experimentally in
Ref.~\cite{Chumak-muMC}. The standard material parameters of Py
used to simulate the dynamics of the magnetization are: saturation
magnetization $\mu_{0}M_{S}=1$~T, exchange stiffness constant $A =
1.3\times10^{-11}$~J/m, and the magnetocrystalline anisotropy is
assumed to be zero. The damping constant in the
Landau-Lifshitz-Gilbert equation is considered to be $\alpha =
0.01$. In order to avoid spin-wave reflection at the ends of the
structure ($x = 0$ and $x = 7.25~\mu$m) the damping boundary
conditions were used: In the boundary areas ($\sim 1 \mu$m on each
side) the parameter $\alpha$ gradually increases more than fifty
times resulting in a strong decay of the spin-wave intensity
towards the boundaries.

In order to excite spin waves within a wide frequency range we
apply a 70~ps-long video pulse to a 300~nm wide microstrip antenna
(see Fig.~\ref{Fig1}). In this case the upper frequency limit
given by the pulse duration is $\sim$13.5~GHz while the antenna
can excite spin-waves with wavenumbers up to $k_\mathrm{max} =
21$~rad/$\mu$m.  The static biasing magnetic field of 500~Oe is
applied in the film plane transverse to its long axis in order to
satisfy the conditions for the excitation and propagation of the
magnetostatic surface spin waves (MSSWs). These waves are known to
have the highest values of the group velocity in magnetic
microstructures and ensure spin-wave free paths of the order of
several microns in Py.

For the understanding of the spin-wave excitation spectra in the
magnonic crystals one need to take into account the distribution
of the internal field shown in Fig.~\ref{Fig1}c and
Fig.~\ref{Fig1}d for both MCs configurations. One can see that the
magnetic field is non-uniform along both the long and short axis
of the crystals. The field variation along the waveguides is not
of crucial importance since it changes mostly near the ends of the
waveguides (see the longitudinal field profiles along shown in
Fig.~\ref{Fig1}e and Fig.~\ref{Fig1}f) where the spin-wave damping
is artificially increased and no spin-wave propagation occurs. In
contrast, the transverse field variation is of high importance:
Even a small deviation from the longitudinal axis of
$0.375$~$\mu$m results in a shift in the field of about 30~Oe.
This means that the internal field creates a spin-wave propagation
channel in the middle of the waveguide \cite{demidov-1}. As is
visible in the figure the magnetic field is practically not
modulated along this channel and the spin wave propagates in a
non-periodic potential. Nevertheless, one can see that the bias
field oscillates closer to the lateral edges of the waveguide (see
dotted curves in Fig.~\ref{Fig1}e-f). This can be considered as a
periodic variation of the effective width of the spin-wave
channel.

\begin{figure}
\includegraphics[width=6 cm]{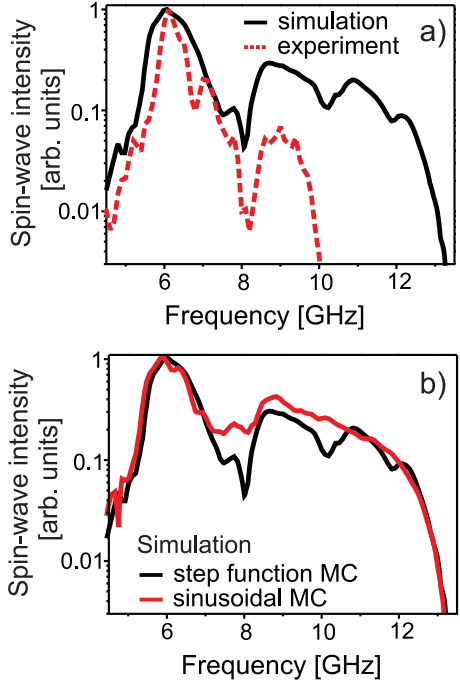}
\caption{\label{Fig2} (Color online) Comparison of spin-wave
transmission characteristics extracted from: a) experiment
\cite{Chumak-muMC} and simulation of MC with rectangular notches
and b) both simulated structures (rectangular and sinusoidal
MCs).}
\end{figure}

In Fig.~\ref{Fig2}a the solid line represents the spin-wave
transmission characteristics extracted from the simulated data for
the notched MC. It was obtained by integrating the spin-wave
intensity along the line drawn across the crystal at $4~\mu$m away
from the antenna (see dotted line in Fig.~\ref{Fig1}a). As it is
expected no spin-wave transmission is allowed below the
ferromagnetic resonance (FMR) frequency $f_\mathrm{FMR}\approx
5$~GHz as well as above the excitation cut-off frequency
$f_\mathrm{max} \equiv f(k_\mathrm{max}) \simeq 13$~GHz.

The spin-wave band gap corresponding to the first Bragg wavenumber
$k_{\Lambda1} = \pi/\Lambda = 3.14$~rad/$\mu$m is visible at $\sim
8$~GHz. The second band gap ($k_{\Lambda2} = 2\pi/\Lambda =
6.28$~rad/$\mu$m) and the third one ($k_{\Lambda3} = 3\pi/\Lambda
= 9.42$~rad/$\mu$m) are not as pronounced but are also clearly
visible around 10.2~GHz and 11.9~GHz, respectively. The
experimentally measured transmission characteristics
\cite{Chumak-muMC} for the same magnonic crystal is shown in
Fig.~\ref{Fig2}a with a dashed line. One can see that the first
band gap is pronounced and has practically the same frequency as
in the simulated data \cite{comment-1}. The higher-order rejection
bands are not visible in the experiment due to limitation of the
wavenumbers that can be excited by the $1~\mu$m-wide antenna:
$k_\mathrm{max} = 6.28$~rad/cm (note that in the simulations we
use a 300~nm-wide antenna). Nevertheless, the results of the
simulations are in good agreement with the experimental ones.
Thus, our approach is trustworthy and can be used for further
studies of MCs properties and characteristics.

The comparison of the simulated transmission characteristics
extracted from both notched and sinusoidal magnonic crystals is
shown in Fig.~\ref{Fig2}b. One sees that the first band gap is
visible for both structures and has exactly the same frequency.
However, the second and the third band gaps do not exist for the
sinusoidal magnonic crystal in contrast to the notched structure.
This brings us to an important conclusion: The MCs transmission
characteristics can be changed qualitatively by designing the MCs
geometry. In addition, one can see that the first band gap is not
so pronounced for the sinusoidal structure. This is in a good
agreement with the results obtained in Ref.~\cite{Chumak-muMC},
where it has been shown that approximately half of the incident
wave energy is reflected back from the notches due to the
geometrical mismatch between the waveguide sections. In the case
of a sinusoidal magnonic crystal this reflection vanishes which
results in a smaller reflection parameter and a weaker rejection
efficiency, respectively.

In general, several scattering mechanisms can contribute to the
formation of the band gaps in the in-plane shaped magnonic
crystals: (1) the reflection caused by the variation of the
transverse spin-wave wavenumber (this wavenumber is inverse to the
effective width of the spin-wave waveguide determined by the
parameters of the spin-wave propagation
channel)~\cite{Chumak-muMC}, (2) the scattering connected with the
transformation of the spin-wave wavevector in the non-uniform
internal magnetic field~\cite{Kostylev_scattering}, and (3) the
spin-wave reflection from the notches due to the geometrical
mismatch (this mechanism is pronounced only for the notched MC).

Since in our case there is a relatively smooth internal field
distribution inside of the spin-wave channel, the first mechanism
appears to be dominant. We have estimated its role in the
formation of the multiple rejection bands. In the approach of the
uniform internal field, we can express the variation of the width
in terms of wavenumbers as follows: The total wavevector of the
propagating wave is given by:
$|\overrightarrow{k}|=\sqrt{k_{L}^{2}+k_{W}^{2} + k_{T}^{2}}$, in
which $k_{L}$ represents the longitudinal wavevector,
$k_{W}=(n\pi/w)$ and $k_{T} = (m\pi/t)$ are the transverse
quantized wavenumbers due to the finite width ($w$) and finite
thickness ($t$), respectively. The integer parameters $n$ and $m$
represent the width and the thickness mode numbers. In our case
$k_{T}=0$ is due to the small Py thickness and the uniform
distribution of the dynamic magnetization on the film thickness.
Moreover, $k_{W}$ becomes a function of the coordinate along the
structure, and for sinusoidal and step function modulation we will
have the corresponding relations:

\begin{equation}\label{wvsin}
k_{w,\textmd{sinus}}^2(x) = \left(\frac{n\pi}{w}\right)^2
\left[1-2\frac{2\delta}{w}\sin\left(\frac{2\pi}{\Lambda}x\right) +
o \left(\frac{2\delta}{w}\right) \right]
\end{equation}

\begin{equation}\label{wvstep} k_{w,\textmd{step}}^2(x) =
\left(\frac{n\pi}{w}\right)^2 \left[1 - 2
\frac{2\delta}{w}\sum_{p=1}^{\infty} \left(\frac{1-(-1)^p}{\pi
p}\right) \times \sin \left(\frac{2p\pi}{\Lambda}x\right) + o
\left(\frac{2\delta}{w}\right) \right]
\end{equation}
where $\delta$ is the modulation amplitude, $x$ is the spatial
coordinate along the structure, $\Lambda$ is the period of the
modulation, and $p$ is the number of space harmonics in the
Fourier series. The term $o(2\delta/w)$ represents the higher
space harmonics of the parameter $2\delta/w$ with amplitudes
orders smaller than the first term. One can observe that both
equations predict multiple rejection bands. However, the
amplitudes given by the higher order terms $o(2\delta/w)$ are
negligible in comparison with the ones given by the first term.
This analytical prediction has been successfully proved by
performing an additional simulation in which we considered a
sinusoidally shaped waveguide with the maximal width variation of
0.75~$\mu$m ($50\%$ increase in the modulation depth in comparison
with the previously shown data). The simulation has shown a single
band gap in the spin-wave transmission characteristics.

\begin{figure}
\includegraphics[width=8.5 cm]{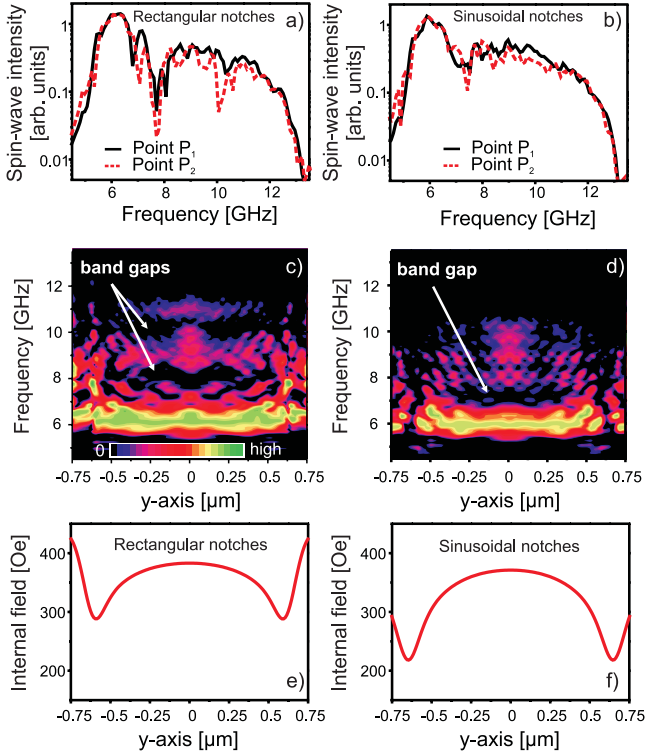}
\caption{\label{Fig3} (Color online) a) and b) Transmission
characteristics extracted locally in the points $P_{1}$ (in the
center) and $P_{2}$ (at $y = 375$~nm) (see Fig.~\ref{Fig1}a and
Fig.~\ref{Fig1}b) inside the MCs. Transmission characteristics as
a function of the probing position across the spin-waves
propagation channel for the c) step function and d) sinusoidal
structures at a distance $x= 4~\mu$m from the antenna. e) and f)
$H_{y}$ component of the magnetic internal field across the
simulated structures.}
\end{figure}

A deeper analysis reveals the dependence of the frequency and the
depth of the rejection bands on the distance from the excitation
antenna. It has been found that the band gaps appear already after
the spin wave has passed two structural periods of the magnonic
crystal. Furthermore, the simulation data shows that the rejection
band depth depends also on the probing position across the MCs
width. Figures~\ref{Fig3}a and ~\ref{Fig3}b present the
transmission characteristics extracted in two points: $P_{1}$ in
the center of the structure (see Fig.~\ref{Fig1}a), and $P_{2}$ at
half the distance between the center and the edge of the spin-wave
propagation channel ($y = 375$~nm). One may note that the depth of
the rejection bands is increased for both simulated structures.
The dependence of the magnitude of the precessing magnetization
component (or spin-wave amplitude) as a function of the position
across the MC structures are shown in Fig.~\ref{Fig3}c and
Fig.~\ref{Fig3}d. Several features are clearly visible: (1) the
band gaps are more pronounced in the regions close to the edges of
the waveguide rather than in the center, (2) the band gap
frequency is slightly varied with the position, and (3) the
absence of the rejection bands just near the waveguide edges. To
understand the first feature one should consider that besides the
modulation of the transversal wavevector, there is still a
variation of the internal field. This variation is more pronounced
closer to the notches than in the center of the magnonic crystal.
Therefore, the spin wave which propagates close to the MCs edge is
more sensitive to the modulation of the internal field which
results in an increase of the rejection efficiency and a more
pronounced band gap. The second feature is related to the decrease
of the internal field towards the notches (see Fig.~\ref{Fig3}e
and ~\ref{Fig3}f). The change of the internal field produces a
shift of the spin-wave dispersion characteristics triggering the
shift of the band gap frequency. Moreover, the temporal evolution
of the spatial distribution of the magnetization shows a
complicated behavior at the edge of the structure. Standing waves
formed inside the anti-notches are mixed with secondary
higher-order width modes generated due to the scattering of the
originally excited one at the edges \cite{KimLee}. As a
consequence, the band gaps are vanishing at the edges of the
spin-wave propagation channel (the third feature).

In conclusion, we performed a micromagnetic study of the spin-wave
transmission in in-plane shaped microstructured magnonic crystals.
Two kinds of spin-wave Permalloy waveguides magnetized across
their longitudinal axes were investigated: with notched and
sinusoidal edge profiles. In both cases spin waves propagate
through a magnetically induced channel of varying effective width.
It has been found that the waveguide width variation leads to the
modulation of the spin-wave wavevectors, and consequently to the
formation of frequency rejection bands in the spin-wave
transmission characteristics. Both the band gap frequency and its
depth depend strongly on the probing position inside the magnonic
crystal due the non-uniform distribution of the internal magnetic
field. The band gap properties including their number can be
predefined by a proper choice of the magnonic crystal geometry.

Financial support by the DFG SE-1771/1-2 is gratefully acknowledged.


\begin{thebibliography}{20}
\bibitem{comment-1} To note that the spin-wave intensity was recorded in the experiment after 8 structure periods while in the simulation after 4 periods. Hence, one may observe a difference in the depth and the width of the first rejection band between the experimental data and simulation, and also a slight shift in frequency.
\end{thebibliography}
\end{document}